\documentstyle[12pt]{article}\textheight 220mm\textwidth 160mm
\newcommand{\bi}{\bibitem}
\newcommand{\bea}{\begin{equation}}
\newcommand{\eea}{\end{equation}}
\newcommand{\be}{\begin{eqnarray}}
\newcommand{\ee}{\end{eqnarray}}
\newcommand{\nn}{\nonumber}
\catcode`\@=11
\def\lsim{\mathrel{\mathpalette\@versim<}}
\def\gsim{\mathrel{\mathpalette\@versim>}}
\def\@versim#1#2{\vcenter{\offinterlineskip
\ialign{$\m@th#1\hfil##\hfil$\crcr#2\crcr\sim\crcr } }}
\catcode`\@=12
\begin{document}
\pagestyle{empty}
\hspace*{11cm}\vspace{-3mm} DFTT 29/96\\ 
\hspace*{11.7cm}\vspace{-3mm}KANAZAWA-96-09\\
\hspace*{11.7cm}June 1996

\begin{center}
{\Large\bf  Unification of Gauge and Yukawa Couplings
without Symmetry$\ ^{*}$}
\end{center} 

\begin{center}{\sc Jisuke Kubo}$\ ^{(1),**}$, 
{\sc Myriam
Mondrag{\' o}n}$\ ^{(2)}$, \vspace{-1mm}\\
{\sc Marek Olechowski}$\ ^{(3),\dag}$ and 
{\sc George Zoupanos}$\ ^{(4),\ddag}$  
\end{center}
\begin{center}
{\em $\ ^{(1)}$ 
Physics Department, \vspace{-2mm} Faculty of Science, \\
Kanazawa \vspace{-2mm} University,
Kanazawa 920-11, Japan } \\
{\em $\ ^{(2)}$ Instituto de F{\' \i}sica,  \vspace{-2mm} UNAM,
Apdo. Postal 20-364,
M{\' e}xico 01000 D.F., M{\' e}xico}\\
{\em $\ ^{(3)}$ INFN Sezione di Torino and 
Dipartamento di Fisica Teorica,  \vspace{-2mm}\\ Universit{\` a} 
di Torino
      Via P. Giuria 1, 10125 Turin, Italy}\vspace{-2mm}\\
{\em $\ ^{(4)}$  Physics Department, \vspace{-2mm} National
Technical University, \\ GR-157 80 Zografou, Athens, Greece}  
  \end{center}

\begin{center}
{\sc\large Abstract}
\end{center}

\noindent
A natural gradual 
extension of the idea of Grand Unification is 
\vspace{-2mm} to attempt to relate 
the gauge and Yukawa
couplings;  Gauge-Yukawa
Unification (GYU). 
\vspace{-2mm} However, within the framework of renormalizable field
theories, \vspace{-2mm} there exists no realistic symmetry that leads to
a GYU. Here we propose an approach to \vspace{-2mm} GYU which is based
on the  principle of the reduction of couplings
and \vspace{-2mm} finiteness in supersymmetric Grand Unified Theories.
We elucidate how 
 the observed \vspace{-2mm} top-bottom mass hierarchy
can be  explained 
in terms of supersymmetric GYU by considering 
an example \vspace{-2mm} of the $SU(5)$ Finite Unified Theory.
 It is expected that, when more accurate  \vspace{-2mm} measurements of
the top and bottom quark masses are available,  it will be possible to
discriminate among the various GYU models.

\vspace*{0.5cm}
\footnoterule
\vspace*{2mm}
\noindent
$^{*}$ Presented by J.K. at the $5^{\rm th}$ \vspace{-3mm} Hellenic
School and Workshops on Elementary Particle Physics,
September 3-24, 1995, Corfu, \vspace{-3mm}
 to appear in the proceedings.\\
$^{**}$Partially supported  by the Grants-in-Aid
for \vspace{-3mm} Scientific Research  from the Ministry of
Education, Science 
and Culture \vspace{-3mm}  (No. 40211213).\\
\noindent
$ ^{\dag}$On leave of absence  from
the Institute of \vspace{-3mm} Theoretical Physics,
Warsaw University,
ul. Hoza 69, 00-681 Warsaw, \vspace{-3mm} Poland.
Partially supported  by the Polish Committee for Scientific
Research. \vspace{-3mm}\\
\noindent
$ ^{\ddag}$
Partially supported \vspace{-3mm} by C.E.C. projects, SC1-CT91-0729;
CHRX-CT93-0319.

\newpage
\pagestyle{plain}

\section{Introduction}

The traditional way to reduce the independent parameters of a theory
is the introduction of a symmetry.  Grand Unified Theories (GUTs)
\cite {pati1,gut1,GUTS} are
representative examples of such attempts.  For instance, the minimal
$SU(5)$ reduces the gauge couplings of the 
Standard Model (SM) by one and gives us a
testable prediction for one of them.  In fact, LEP data \cite{abf,lec1} 
seem to
suggest that the $N=1$ global supersymmetry
\cite{sakai1,lec2} 
should  be required in addition to make the prediction viable. 
GUTs also relate Yukawa couplings among themselves, which can
 lead to  predictions for the parameters of the SM. 
The
prediction of the ratio $m_{\tau}/m_b$ \cite{begn} in the minimal
$SU(5)$ was an 
example of a successful reduction of the independent parameters of
this sector. On the other hand, requiring more symmetry
(e.g., $SO(10),~E_6,~E_7,~E_8$) does not necessarily lead to more
predictions for the SM parameters, due to the presence of new degrees
of freedom, various ways and channels of breaking the symmetry, etc.  An
extreme case from this point of view are superstrings, which have huge
symmetries, but no real predictions for the SM parameters.

In a series of papers \cite{kmz}--\cite{kmoz2},
we have proposed
that a natural gradual 
extension of the GUT ideas, which preserves their successes and
enhances the predictions, is to attempt to relate 
the gauge and Yukawa
couplings, or in other words, to achieve Gauge-Yukawa
Unification (GYU).  
 Searching for a symmetry that could provide such a
unification, one is led to introduce a symmetry that relates fields
with different spins, i.e., supersymmetry and in particular
$N=2$ supersymmetry \cite{f79}.  
Unfortunately, $N=2$ supersymmetric theories have
serious phenomenological problems due to light mirror fermions.
Needless to say that there exists GYU 
in superstrings, too \cite{lec3,lec4}.

In the following we would like to emphasize an alternative way to
achieve unification of couplings, which is based on the fact that
within the framework of a renormalizable field theory, one can find
renormalization group (RG) invariant relations among parameters,
that can improve the calculability and the predictive power of a
theory.  In our recent studies 
\cite{kmz}--\cite{kmoz2}, we have
considered the GYU 
which is based on the principles of reduction of couplings
\cite{chang1}--\cite{andrianov1} and
finiteness \cite{PW}--\cite{LZ}.  These
principles, which are formulated in 
perturbation theory, are not explicit symmetry principles, although
they might imply symmetries.  The former principle is based on the
existence of RG invariant relations among couplings, which preserve
perturbative renormalizability.  Similarly, the latter one is based on
the fact that it is possible to find RG invariant relations among couplings
that keep finiteness in perturbation theory, even to all orders.
Applying these principles one can relate the gauge and Yukawa
couplings without introducing necessarily a symmetry, nevertheless
improving the predictive power of 
a model. Concerning recent related studies, we would like to
emphasize that our approach to GYU for asymptotically non-free
theories \cite{kmtz,kmtz2} covers work done by other authors
\cite{lanross}, though the underlying 
idea might be different.

In the next section we begin by illustrating the idea of reduction
of couplings, and in section 3 we consider
a Finite Unified Theory (FUT) based on $SU(5)$--one
of the successful Gauge-Yukawa Unified theories--which, moreover,
is attracting a renewed interest because of duality in
supersymmetric field theories \cite{seiberg,leigh}.

\section{Reduction of couplings}

To illustrate the idea of the reduction of couplings, we consider
a  theory containing two scalar fields
 $\phi_I ~,~I=1,2$.
The renormalizable
Lagrangian, which has two parities
$ \phi_I \to -\phi_I $,  is given by
\be
{\cal L} &=& \frac{1}{2}\sum_{I=1,2}~(~\partial_{\mu}
\phi_I\partial^{\mu}\phi_I-m^{2}_{I}~\phi_{I}^{2}~)
-\frac{g_1}{4!}~\phi_{1}^{4}-
\frac{g_0}{4}~\phi_{1}^{2}\phi_{2}^{2}
-\frac{g_2}{4!}~\phi_{2}^{4}~.
\ee
The theory defined by this Lagrangian has 
originally three dimensionless couplings
$g_i~,~i=0,1,2$ and two dimensionful parameters $m_1$ and $m_2$,
and we would like to consider the reduction in these numbers.

To this end, we first compute one-loop diagrams 
in $4-2\epsilon$ dimensions and employ
the minimal subtraction (MS) scheme for renormalization.
One finds in this order
\be
g_{0}^{(0)} &=&\mu^{2\epsilon}~[~
g_0+\frac{1}{\epsilon}
\frac{1}{16\pi^2}~(~g_{0}^{2}+\frac{1}{4}g_{1}g_0
+\frac{1}{4}g_{2}g_0~)~]~,\\
g_{i}^{(0)} &=&\mu^{2\epsilon}~[~
g_i+\frac{1}{\epsilon}
\frac{1}{16\pi^2}~(\frac{3}{4})~(~g_{0}^{2}+g_{i}^{2}~)~]~(i=1,2)~,\\
(m_{1}^{(0)})^2 &=&m_{1}^{2}+[~\frac{1}{\epsilon}
\frac{1}{16\pi^2}~(\frac{1}{2})~(g_{1}m_{1}^{2}
+g_{0}m_{2}^{2}~)~]~,\\
(m_{2}^{(0)})^2 &=&m_2+[~\frac{1}{\epsilon}
\frac{1}{16\pi^2}~(\frac{1}{2})~(g_{2}m_{2}^{2}
+g_{0}m_{1}^{2}~)~]~,~
\ee
where $\mu$ is the 't Hooft renormalization scale,
and $g^{(0)}$'s and
$m^{(0)}$'s stand for the bare couplings and  masses.
To maintain renormalizability of the theory,
it is usually assumed that these five parameters are independent.
There may be, however, exceptional situations.
Obviously, in the presence of the $O(2)$ symmetry,
we have $m_1=m_2$ and $g_1=g_2=3g_0$ so that
only one dimensionless and one massive parameter are
independent. This is true to all orders in perturbation theory,
because the $O(2)$ symmetry is anomaly-free in the present case.

Are there other possibilities?
To answer this question at one-loop order, we assume that
\be
g_i &=& \rho_i\,g_0~,~(i=1,2)~,~m_1=e\,m_2~,
\ee
and insert them into the renormalization eqs. (2)--(5).
One finds that (under the assumption that $m_{1}^{2}~,~m_{2}^{2} > 0$)
there are two 
solutions that are consistent with the one-loop
renormalizability:
\be
\rho_1 &=&\rho_2=3~\mbox{or}~\rho_1 = \rho_2=1
\ee
with $e=1$.
The first one is the symmetric one, but the second one 
is associated with no obvious symmetry.
So, the second one might be an artifact of one-loop order
and could disappear if one goes to higher orders.
It is remarkable that one can check at one-loop order already,
whether the second possibility of reducing the number of parameters
persists in higher orders. We will see it in a moment.

The reduction of couplings
 was originally formulated for a massless theory
on the basis of the Callan-Symanzik equation.
The extension to theories with massive parameters
is not straightforward if one wants to keep
the generality and the rigor
on the same level as for the massless case;
one has 
to fulfill a set of requirements coming from
the renormalization group
equations,  the  Callan-Symanzik equations, etc.,
along with the normalization
conditions imposed on irreducible Green's functions.
There have been some progresses in this direction \cite{maison}. 
Here we would like  to present 
the idea of the reduction of dimensionless couplings.
As we have done in the example above, we 
 assume, to make the method transparent,  that
 the MS scheme has been
employed so that all the  RG functions such as $\beta$ functions
 depend only on dimensionless couplings.
Then we would like to investigate whether a solution like
eq. (7), which is not a consequence of a symmetry, 
persists to higher orders in
perturbation theory. 

To be general, we consider  a massless renormalizable theory
which contain a set of $(N+1)$ dimensionless couplings. 
The renormalized irreducible Green's function 
in the MS scheme satisfies the RG equation
\be
0 &=&[~\mu\frac{\partial}{\partial \mu}+
~\beta_i\,\frac{\partial}{\partial g_i}+
~\Phi_I \gamma^{\phi I}_{~~~J}
\frac{\delta}{\delta \Phi_J}
~]~\Gamma (~{\bf \Phi},g_0,g_1,\dots,g_N,\mu~)~,
\ee
where ${\bf \Phi}$ stands for a set of fields, $\beta$'s
for the $\beta$ functions and $\gamma$
for the $\gamma$ functions.
We then ask ourselves whether the reduction of parameters, i.e.,
\be
g_i &=&g_i(g)~,~(i=1,\dots,N)~,~g\equiv g_0
\ee
is consistent with the RG equation
\be
0 &=&[~\mu\frac{\partial}{\partial \mu}+
~\beta_g\,\frac{\partial}{\partial g}+
~\Phi_I  \gamma^{\phi I}_{~~~J}\frac{\delta}{\delta \Phi_J}
~]~\Gamma (~{\bf \Phi},g,g_1(g),\dots,\mu~)~,
\ee
where $g$ is called the primary coupling.
We find  that the following set of
equations has to be satisfied:
\be
\beta_g =\beta_0 &,& 
\beta_g\,\frac{d g_i}{d g} = \beta_i~,~(i\neq 0)~,
\ee
which are called the reduction equations \cite{zim1}.

The bare quantities are given by
\be
\Phi^{(0)}_{I} &=&
 \mu^{k_{ I}\epsilon}Z^{\phi~J}_{I} (g) \Phi_J~,~
g_{i}^{(0)} = \mu^{k_{i}\epsilon} Z^{g~j}_{i}(g) g_j (g)~.
\ee
The renormalization constants  above are those which
 are first computed in
the original theory and then rewritten by means of 
eq. (9), and the $k$'s are introduced to match the dimension
in $(4-2 \epsilon)$ dimensions.
Therefore, the requirements for the reduced theory to be perturbative
renormalizable means that the functions $g_i (g)$
should have a power series
expansion in the primary coupling $g$. That is,
\be
g_{i}(g) &=& g\,\sum_{n=0}^{\infty} \rho_{i}^{(n)} 
g^{n}~(i\neq 0)~.
\ee
REcalling ourselves that $\beta$'s and $\gamma$'s are also
a power series and assuming that the expansion coefficients
with $n \leq n_0$ are determined, 
 we insert the power series ansatz (13) into the
 reduction equations (11).
 One finds that to obtain the $(n_0+1)^{\mbox{th}}$
order coefficients, we have to solve a linear system 
of equations with $N$
unknown quantities, where its coefficients are given
by the lowest order quantities in the reduction procedure.
This is the reason why one can investigate at the lowest order,
whether the linear system in  $(n_0+1)^{\mbox{th}}$
 order  can be uniquely solved.

For our example of a $\phi^4$ theory, one finds 
\be
\beta_0 &=& \mu \frac{d g_0}{d \mu} = \frac{1}{16\pi^2}
(4g_0^2+g_1 g_0+g_2 g_0)+\dots ~,\\
\beta_i &=& \mu \frac{d g_i}{d \mu} = \frac{3}{16\pi^2}
( g_0^2+g_{i}^{2})+\dots ~,~(i=1,2)~,
\ee
where $\dots$ indicates higher order terms.
The power series ansatz for the present case
 takes the form
\be
g_{i}(g) &=& g\,(~\rho_{i}^{(0)}+\sum_{n=1}^{\infty} \rho_{i}^{(n)}
g^{n}~)~,~(i=1,2)~,
\ee
 where 
\be
\rho_{1}^{(0)}=\rho_{2}^{(0)}=3~\mbox{or}~1~.
\ee
As described above, we insert them into the 
corresponding reduction equations  and assume that $\rho_{i}^{(n)}$
with $n \leq n_0$ are determined already. Collecting terms
of $O(g^{n_0+3})$,
we find that
\be
 & &\left( \begin{array}{ll}
(n_0+2)(4+\rho_{1}^{(0)}+\rho_{2}^{(0)})-5\rho_{1}^{(0)}
&~~~~~\rho_{1}^{(0)}\\
~~~~~\rho_{2}^{(0)}
& (n_0+2)(4+\rho_{1}^{(0)}
+\rho_{2}^{(0)})-5\rho_{2}^{(0)}\\ 
 \end{array} \right) \nn\\
& &\times \left( \begin{array}{l}
\rho_{1}^{(n_0+1)}\\
 \rho_{2}^{(n_0+1)}\\
 \end{array} \right) =
 \mbox{known quantities}~.
\ee
Since the matrix on the l.h. side of eq. (18)
is regular,
 we conclude that $\rho_{i}^{(n_0+1)}$
can be uniquely determined.
That is, the power series (13) exists uniquely.

Moreover, it is possible \cite{zim1} to find 
a  reparametrization of couplings in such a way that
$\rho_{i}^{(n)}$ for all $n > 0$ exactly vanish.
In fact, this theory corresponds to \cite{kazakov1}
$$
{\cal L} =\sum_{I=+,-}~(~ \frac{1}{2}\partial_{\mu}
\phi_I\partial^{\mu}\phi_I
-\frac{g_0}{6}~\phi_{I}^{4}~)~~,~~\phi_{+(-)}=
\frac{1}{\sqrt{2}}(\phi_1+(-) \phi_2)~.$$

\section{Finite Unified Model Based on $SU(5)$}

As a realistic example for the reduction of couplings,
we consider 
a Finite Unified Model Based on $SU(5)$.
From the classification of
theories with vanishing one-loop 
$\beta$ function for the gauge coupling
\cite{HPS}, one can see that
using $SU(5)$ as gauge group there
exist only two candidate models which can 
accommodate three fermion
generations. These models contain the chiral supermutiplets
${\bf 5}~,~\overline{\bf 5}~,~{\bf 10}~,
~\overline{\bf 5}~,~{\bf 24}$
with the multiplicities $(6,9,4,1,0)$ and
 $(4,7,3,0,1)$, respectively.
Only the second one contains a ${\bf 24}$-plet which can be used
for spontaneous symmetry breaking (SSB) of $SU(5)$ down
to $SU(3)\times SU(2) \times U(1)$. (For the first model
one has to incorporate another way, such as the Wilson flux
breaking to achieve the desired SSB of $SU(5)$.)
Therefore,  we would like to concentrate only on the second model.

To simplify the situation, we neglect the intergenerational
mixing among the lepton and quark supermultiplets and consider
the following $SU(5)$ invariant cubic
superpotential for the (second)
model:
\be
W &=& \sum_{i=1}^{3}\,[~\frac{1}{2}g_{i}^{u}
\,{\bf 10}_i
{\bf 10}_i H_{i}+
\sqrt{2}g_{i}^{d}\,{\bf 10}_i \overline{\bf 5}_{i}\,
\overline{H}_{i}~] \nn\\
 & & +\sum_{\alpha=1}^{4}g_{\alpha}^{f}\,H_{\alpha}\, 
{\bf 24}\,\overline{H}_{\alpha}+
\frac{g^{\lambda}}{3}\,({\bf 24})^3~,
\ee
where the ${\bf 10}_{i}$'s
and $\overline{\bf 5}_{i}$'s are the usual
three generations, and the four
$({\bf 5}+ \overline{\bf 5})$ Higgses are denoted by
 $H_{\alpha}~,~\overline{H}_{\alpha} $.
The superpotential is not the most general one, but
by virtue of the non-renormalization theorem,
this does not contradict the philosophy of 
the coupling unification by the reduction 
method. (A RG invariant fine tuning is a solution
of the reduction equation \footnote{In the case at hand,
however, one can
find a discrete symmetry that can be imposed
on the most general cubic superpotential to arrive at the
non-intergenerational mixing \cite{kmz}.} ).
Given the superpotential $W$,
we  can  compute the $\beta$ functions of the model. 
We denote the gauge coupling  by $g$
(with the vanishing one-loop $\beta$ function), and 
our  normalization of the $\beta$ functions
is as usual, i.e., 
$d g_{i}/d \ln \mu ~=~
\beta^{(1)}_{i}/16 \pi^2+O(g^5)$,
where $\mu$ is the renormalization
scale. We find: 
\be
\beta^{(1)}_{g} &=& 0~,\nn\\
\beta^{u(1)}_{i} &=& \frac{1}{16\pi^2}\,
[\,-\frac{96}{5}\,g^2+
9\,(g_{i}^{u})^{2}+
\frac{24}{5}\,(g^{f}_{i})^{2}+
4\,(g_{i}^{d})^{2}
\,]\,g_{i}^{u}~,\nn\\
\beta^{d(1)}_{i} &=& \frac{1}{16\pi^2}\,
[\,-\frac{84}{5}\,g^2+
3\,(g_{i}^{u})^{2}
+\frac{24}{5}\,(g^{f}_{i})^{2}+
10\,(g_{i}^{d})^{2}\,]\,g_{i}^{d}~,\\
\beta^{\lambda(1)} &=& \frac{1}{16\pi^2}\,
[\,-30\,g^2+\frac{63}{5}\,(g^{\lambda})^2+
3\,\,\sum_{\alpha =1}^{4}(g_{ \alpha}^{f})^{2}
\,]\,g^{\lambda}~,\nn\\
\beta^{f(1)}_{\alpha} &=& \frac{1}{16\pi^2}\,
[\,-\frac{98}{5}\,g^2+3\,(g_{i}^{u})^{2}\delta_{i\alpha}
+4\,(g_{i}^{d})^{2}\delta_{i\alpha}
+\frac{48}{5}\,(g^{f}_{\alpha})^{2}
+\sum_{\beta=1}^{4}(g_{\beta}^{f})^{2}+\frac{21}{5}\,(g^{\lambda})^{2}
\,]\,g_{\alpha}^{f}~.\nn
\ee
We then regard the gauge coupling $g$ as the primary
coupling and solve the reduction equations (11) with
the power series ansatz. One finds that the power series,
\be
(g_{i}^{u})^2 &=&\frac{8}{5}g^2+\dots~,
~(g_{i}^{d})^2 =\frac{6}{5}g^2+\dots~,~
(g^{\lambda})^2=\frac{15}{7}g^2+\dots~,\nn\\
(g^{f}_{4})^2 &=& g^2~,~(g^{f}_{\alpha})^2=0+\dots~~(\alpha=1,2,3)~,
\ee
exists uniquely,
where $\dots$ indicates higher order terms and
all the other couplings have to vanish.
As we have done in the previous section, we can easily 
verify that the higher order terms can be uniquely
computed.
 Consequently, all the one-loop $\beta$ functions of the theory vanish.
Moreover, all the one-loop anomalous dimensions for the chiral
supermultiplets,
\be
\gamma^{(1)}_{{\bf 10}i} &=& \frac{1}{16\pi^2}\,
[\,-\frac{36}{5}\,g^2+
3\,(g_{i}^{u})^{2}+
2\,(g_{i}^{d})^{2}
\,]~,\nn\\
\gamma^{(1)}_{\overline{{\bf 5}}i} &=& \frac{1}{16\pi^2}\,
[\,-\frac{24}{5}\,g^2+
4\,(g_{i}^{d})^{2}\,]~,\nn\\
\gamma^{(1)}_{H_{\alpha}} &=& \frac{1}{16\pi^2}\,
[\,-24\,g^2+
3\,(g_{i}^{u})^{2}\delta_{i\alpha}+
\frac{24}{5}(g_{\alpha}^{f})^2\,]~,\\
\gamma^{(1)}_{\overline{H}_{\alpha}} &=& \frac{1}{16\pi^2}\,
[\,-24\,g^2+
4\,(g_{i}^{d})^{2}\delta_{i\alpha}+
\frac{24}{5}(g_{\alpha}^{f})^2\,]~,\nn\\
\gamma^{(1)}_{{\bf 24}} &=& \frac{1}{16\pi^2}\,
[\,-\frac{10}{5}\,g^2+
+\sum_{\alpha=1}^{4}(g_{\alpha}^{f})^{2}
+\frac{21}{5}\,(g^{\lambda})^{2}\,]~,\nn
\ee
also vanish in the reduced system.
A very interesting result is that these conditions  are
necessary and sufficient for finiteness at
the two-loop level \cite{PW}.

A natural question is what happens in higher loops.
Interestingly,
 there exists a powerful theorem \cite{LPS}
which provides the necessary and sufficient conditions for
finiteness to all loops.
The theorem makes heavy use of the non-renormalization
property of the supercurrent anomaly \cite{pisi}.
In fact, the  finiteness theorem can be formulated in terms of
one-loop quantities, and it states
 for supersymmetry gauge theories, the necessary
and sufficient conditions for $\beta_{g}$ and $\gamma$'s to
vanish to all orders are \cite{LPS}: \newline
(a) The validity of the one-loop finiteness conditions, i.e.,
$\beta_{g}^{(1)}=\gamma^{(1)\prime}s =0$.
\newline
(b) The reduction equation (11) admit a unique power series
solution. 
\newline
Since the solution (22)
can be extended to a unique power series in $g$, the reduced
theory (which has a single coupling $g$)  has 
$\beta$ and $\gamma$ functions
vanishing to all orders.
In this way, the Gauge-Yukawa Unification 
is achieved \footnote{There is an alternative way to
find finite theories, which has been found in connection
to duality in supersymmetric theories \cite{leigh}.}.

In most of the previous studies of
the present model \cite{model1,model}, however,
the complete reduction of the Yukawa couplings,
which is necessary for all-order-finiteness,
was ignored.  They have used the freedom
offered by the degeneracy in the one- and two-loop
approximations in order to make
specific ans{\" a}tze that could lead to phenomenologically acceptable
predictions.
In the above model, we found a diagonal solution for the Yukawa
couplings, with each family coupled to a different Higgs.
However, we may use the fact that mass terms
do not influence the RG functions in a certain
class of renormalization schemes, and introduce
appropriate mass terms that permit us to perform a rotation in the Higgs
sector such that only one pair of Higgs doublets, coupled to
the third family, remains light and acquires a
non-vanishing VEV \cite{model}. 
Note that the effective coupling of the Higgs doublets
to the first family after
the rotation is very small avoiding in this way a potential problem
with the proton lifetime \cite{proton}.
Thus, effectively,
we have at low energies the Minimal Supersymmetric Standard Model
(MSSM) with
only one pair of Higgs doublets
satisfying the boudary conditions at $M_{\rm GUT}$
\be
 g_{t}^{2}&  = &\frac{8}{5} g^2+O(g^4)~,~
 g_{b}^{2}=g_{\tau}^{2}=\frac{6}{5} g^2+O(g^4)~,
\ee
where $g_i$ ($i=t, b, \tau$) are the top, bottom
and tau Yukawa couplings
of the MSSM, and the other Yukawa couplings 
should be regarded as free.

Adding soft
breaking terms (which are supposed not to influence the
$\beta$ functions beyond $M_{\rm GUT}$),
we can obtain supersymmetry breaking.
The conditions on the soft breaking terms to preserve
one-loop finiteness have been given already some time ago
\cite{soft}. 
Recently, the same problem
in higher orders has been addressed \cite{jj}.
It is an open problem whether there exists a suitable set of conditions
on the soft terms for all-loop finiteness.

\section{Predictions of Low Energy Parameters}
Since the $SU(5)$ symmetry is spontaneously broken
below $M_{\rm GUT}$, the finiteness conditions 
do not restrict the renormalization property at low energies, and
all it remains is a boundary condition on the
gauge and Yukawa couplings at $M_{\rm GUT}$, i.e., eq. (23).
So we examine the evolution of these couplings according
to their renormalization group equations at two-loop with
this boundary condition.

Below $M_{\rm GUT}$ their evolution is assumed to be
governed by the MSSM. We further assume a unique threshold
$M_{\rm SUSY}$ for all superpartners of the MSSM so that
below $M_{\rm SUSY}$ the SM is the correct effective theory.
We  recall that
$\tan\beta$ is usually determined in the Higgs sector, which however
strongly depends on the supersymmetry breaking terms.
Here we avoid this by using the tau mass $M_{\tau}$ 
as input \footnote{This means that we partly fix the Higgs sector
indirectly.}.
That is, assuming that
\be
M_Z \ll M_{t} \ll M_{\rm SUSY}~,
\ee
we require the matching condition at $M_{\rm SUSY}$ \cite{barger},
\be
\alpha_{t}^{\rm SM} 
&=&\alpha_{t}\,\sin^2 \beta~,~
\alpha_{b}^{\rm SM}
~ =~ \alpha_{b}\,\cos^2 \beta~,
~\alpha_{\tau}^{\rm SM}
~=~\alpha_{\tau}\,\cos^2 \beta~,\nn\\
\alpha_{\lambda}&=&
\frac{1}{4}(\frac{3}{5}\alpha_{1}
+\alpha_2)\,\cos^2 2\beta~,
\ee
to be satisfied \footnote{There are MSSM threshold corrections
to this matching condition \cite{hall1,wright1}, 
which will be discussed
later.}, 
where $\alpha_{i}^{\rm SM}~(i=t,b,\tau)$ are
the SM Yukawa couplings and $\alpha_{\lambda}$ is the Higgs coupling.
 This is our definition of $\tan\beta$, and eq. (25)
 fixes $\tan\beta$, because with a given set of the input
parameters \cite{pdg}, 
\be
M_{\tau} &=&1.777 ~\mbox{GeV}~,~M_Z=91.188 ~\mbox{GeV}~,
\ee
with \cite{pokorski1}
\be
\alpha_{\rm EM}^{-1}(M_{Z})&=&127.9
+\frac{8}{9\pi}\,\log\frac{M_t}{M_Z} ~,\nn\\
\sin^{2} \theta_{\rm W}(M_{Z})&=&0.2319
-3.03\times 10^{-5}T-8.4\times 10^{-8}T^2~,\\
T &= &M_t /[\mbox{GeV}] -165~,\nn
\ee
the matching condition (25) and the GYU
boundary condition at $M_{\rm GUT}$ (23) can be satisfied only for a specific
value of $\tan\beta$. Here  $M_{\tau},M_t, M_Z$
are pole masses, and the couplings are defined in the 
$\overline{\mbox{MS}}$ scheme with six flavors.
The translation from a Yukawa coupling
into the corresponding mass follows according to
\be
m_i&=&\frac{1}{\sqrt{2}}g_i(\mu)\,v(\mu)~,~i=t,b,\tau ~~
\mbox{with} ~~v(M_Z)=246.22~\mbox{GeV}~,
\ee
where $m_i(\mu)$'s are the running masses satisfying
the respective evolution equation at two-loop order.
The pole masses can be calculated from the
running ones of course. For the top mass, we use \cite{barger,hall1}
\be
M_{t} &=&m_{t}(M_t)\,[\,1+
\frac{4}{3}\frac{\alpha_3(M_t)}{\pi}+
10.95\,(\frac{\alpha_3(M_t)}{\pi})^2+k_t 
\frac{\alpha_t(M_t)}{\pi}\,]~,
\ee
where  $k_t \simeq -0.3$ for the range of parameters
we are concerned with in this paper \cite{hall1}.
Note that both sides of eq. (29) contains $M_t$ so that
$M_t$ is defined only implicitly.
Therefore, its determination requires an iteration method.
As for the tau and bottom masses, we assume that
$m_{\tau}(\mu)$ and $m_b(\mu)$ for $\mu \leq M_Z$
satisfy the evolution equation governed by
the $SU(3)_{\rm C}\times U(1)_{\rm EM}$ theory 
with five flavors and use
\be
M_{b}&=&m_b(M_b)\,[\,1+
\frac{4}{3}\frac{\alpha_{3(5{\rm f})}(M_b)}{\pi}+
12.4\,(\frac{\alpha_{3(5{\rm f})}(M_b)}{\pi})^2\,]~,\nn\\
M_{\tau}&=&m_{\tau}(M_{\tau})\,[\,1+
\frac{\alpha_{\rm EM (5f)}(M_{\tau})}{\pi}\,]~,
\ee
where the experimental value of $m_b(M_b)$ is
$(4.1-4.5)$ GeV \cite{pdg}.
The couplings with five flavors entered in eq. (30)
$\alpha_{3(5{\rm f})}$ and $\alpha_{\rm EM (5f)}$
are related to $\alpha_{3}$ and $\alpha_{\rm EM}$ by
\be
\alpha_{3(5{\rm f})}^{-1}(M_Z) &= &\alpha_{3}^{-1}(M_Z)
-\frac{1}{3\pi}\,\ln \frac{M_t}{M_Z} ~,\nn\\
\alpha_{\rm EM (5f)}^{-1}(M_Z) &= & \alpha_{\rm EM}^{-1}(M_Z)-
\frac{8}{9\pi}\,\ln \frac{M_t}{M_Z}~.
\ee
Using the input values given in eqs. (26) and (27), we find
\be
m_{\tau}(M_{\tau})&=&1.771~\mbox{GeV}~,
m_{\tau}(M_{Z})=1.746~\mbox{GeV}~,
\alpha_{\rm EM (5f)}^{-1}(M_{\tau})=133.7~,
\ee
and from eq. (28) we 
obtain 
\be
\alpha_{\tau}^{\rm SM}(M_Z)&=&\frac{g_{\tau}^{2}}{4\pi}
=8.005\times 10^{-6}~,
\ee
which we use as an input parameter instead of $M_{\tau}$.

The matching condition (25)  suffers from the threshold 
corrections coming from the MSSM superpartners:
\be
\alpha_{i}^{\rm SM} \to 
\alpha_{i}^{\rm SM}(1+\Delta_{i}^{\rm SUSY})~,~i=1,2,\dots,\tau~,
\ee
It was shown that these threshold effects to
the  gauge couplings can
be effectively parametrized by just one energy  scale 
\cite{langacker1}. 
Accordingly, we can identify our $M_{\rm SUSY}$ with that defined
in ref.\cite{langacker1}.  This ensures that there are no further
one-loop threshold corrections  to $\alpha_3(M_Z)$ when we
calculate it as a function of  $\alpha_{\rm EM}(M_Z)$ and
$\sin^2\theta_W(M_Z)$.

The same scale $M_{\rm SUSY}$
does not describe  threshold corrections to the Yukawa
couplings,
and they could cause large corrections to the fermion mass
 prediction \cite{hall1,wright1} \footnote{It is
possible to compute the MSSM correction to $M_t$ directly, i.e., 
without constructing an effective theory below $M_{\rm SUSY}$.
In this approach, too,  large corrections have  been reported
\cite{polonsky1}. In the present paper, evidently,  we are following
the effective theory approach as 
e.g. refs. \cite{hall1,wright1}.}.
For $m_b$, for instance, the correction  can be as large as 50\%
for very large values  of $\tan\beta$,
especially in models with radiative 
gauge symmetry breaking and with supersymmetry softly broken by 
the universal breaking terms. As we will see later,
the  $SU(5)$-FUT model predicts (with these corrections suppressed) values
for the bottom quark mass that are 
rather close to the experimentally allowed region 
so that there is room only for small corrections.
Consequently, if we want to break
the $SU(2) \times U(1)$ gauge 
symmetry radiatively, the model favors
non-universal soft breaking terms \cite{borzumati1,lec4}.
It is interesting to note that
the consistency of the finiteness hypothesis
is closely related to the fine structure of supersymmetry breaking
and also to the Higgs sector, because
these superpartner corrections  to $m_b$ can be kept small
for appropriate 
supersymmetric spectrum characterized by very heavy squarks 
and/or small  $\mu_H$ describing the mixing of the two 
Higgs doublets in the superpotential 
\footnote{The solution with small $\mu_H$ 
is favored by the experimental data and cosmological constraints
\cite{borzumati1}. The sign of this correction 
is determined by the relative sign of 
$\mu_H$ and the gluino mass parameter, $M_3$, and is correlated 
with the chargino exchange contribution 
to the $b \to s \gamma$ decay \cite{hall1}. 
The later has the same sign as the Standard Model and the charged 
Higgs contributions when the supersymmetric corrections to $m_b$ are 
negative.}.

To get an idea about the magnitude of the correction, 
we consider
the case that all the superpartners 
have the same mass $M_{\rm SUSY}=500$ GeV with
$M_{\rm SUSY} >> \mu_H$ and $\tan\beta \gsim 50$.
Using  $\Delta$'s 
given in  ref. \cite{wright1}, we find that
the MSSM correction to the $M_t$ prediction
is $\sim -1$ \% for this case.
Comparing with the 
results of \cite{wright1,polonsky1}, 
this may appear to be underestimated.
Note, however, that there is a nontrivial interplay among the 
corrections between the $M_t$ and $M_b$ predictions
for a given GYU boundary condition at $M_{\rm GUT}$
and the fixed pole tau mass, which has not been taken into
account in refs. \cite{wright1,polonsky1}. 
In the following discussion, therefore,
we regard the MSSM threshold correction to
the $M_t$ prediction as unknown and denote it by
\be
\delta^{\rm MSSM} M_t~.
\ee

In table 1 we present
the predictions of $M_t$ and $m_b(M_b)$ for various 
given values of $M_{\rm SUSY}$.

\vspace{1cm}

\begin{center}
\begin{tabular}{|c|c|c|c|c|c|}
\hline
$M_{\rm SUSY}$ [GeV]   &$\alpha_{3}(M_Z)$ &
$\tan \beta$  &  $M_{\rm GUT}$ [GeV] 
 & $m_b (M_{b}) $ [GeV]& $M_{t}$ [GeV]
\\ \hline
$300$ & $0.123 $  &$54.2 $  & $2.08\times 10^{16}$
 & $4.54$  & 183.5\\ \hline
$500$ & $0.122 $  &$54.3 $  & $1.77\times 10^{16}$
 & $4.54$  & 184.0 \\ \hline
$10^3$ & $0.120 $  &$54.4 $  & $1.42\times 10^{16}$
 & $4.54$  & 184.4 \\ \hline
\end{tabular}
\end{center}

\begin{center}
{\bf Table 1}. The predictions 
for different $M_{\rm SUSY}$ for FUT.
\end{center}

\noindent
As we can see from the table,  only negative MSSM
corrections of at most $\sim 10$ \% to $m_b(M_b)$ 
are allowed ( $m_{b}^{\rm exp}(M_b)=
(4.1-4.5)$ GeV), implying that FUT
favors  non-universal soft symmetry breaking terms as announced.
The predicted $M_t$ values are well below the 
infrared value \cite{hill1},
for instance,
$194$ GeV for $M_{\rm SUSY}=500$ GeV, 
so that the $M_t$ prediction
must be sensitive against the change of the boundary condition
(23).

We recall that if one includes
the threshold effects of superheavy particles \cite{threshold,lec5},
the GUT scale $M_{\rm GUT}$ at which $\alpha_1$ and $\alpha_2$
are supposed to meet is related to
the mass of the superheavy
$SU(3)_C $-triplet  Higgs supermultiplets contained 
in $H_{\alpha}$ and $\overline{H}_{\alpha}$.
These  effects  have therefore
influence on the GYU boudary condition (23).
The structure of the threshold effects in FUT
 is involved, but
they are not arbitrary and probably determinable to a certain
extent, because the mixing of the superheavy Higgses
is strongly dictated by the fermion mass matrix of the MSSM.
To bring these threshold effects under control is
challenging.
Here we assume that the magnitude of these effects is
$\sim \pm 4$  GeV in $M_t$
(which is estimated by comparing the minimal GYU model based
on $SU(5)$ \cite{kmoz2}.).
We conclude \cite{kmoz2} that 
\be
M_t &=&(183+\delta^{\rm MSSM} M_t\pm 5) ~~\mbox{GeV}~,
\ee
where the finite corrections coming from the conversion
from the dimensional reduction scheme to
the ordinary $\overline{\mbox{MS}}$ 
in the gauge sector \cite{anton} are included, and
those in the Yukawa sector are included as an uncertainty
of $\sim\pm 1$ GeV.
The MSSM  threshold
correction 
is  denoted $\delta^{\rm MSSM} M_t$
which has been discussed in the previous section.
Comparing the $M_t$ prediction above  with the experimental value \cite{top},
$M_t=(175\pm 9)$ GeV \cite{top}, we see
it is consistent with the
experimental data.

\section{Conclusion}
As a natural extension of the unification of gauge couplings provided by 
all GUTs and the unification of Yukawa couplings, we
have introduced the idea of Gauge-Yukawa
Unification. GYU is a functional relationship among the gauge and
Yukawa couplings provided by some principle.  In our studies GYU has
been achieved by applying the principles of reduction of couplings
and finiteness. 
The consequence of GYU is that 
in the lowest order in perturbation theory
 the gauge and Yukawa couplings above  $M_{\rm GUT}$
are related  in the form
\be
g_i & = &\kappa_i \,g_{\rm GUT}~,~i=1,2,3,e,\cdots,\tau,b,t~,
\ee 
where $g_i~(i=1,\cdots,t)$ stand for the gauge 
and Yukawa couplings, $g_{\rm GUT}$ is the unified coupling,
and
we have neglected  the Cabibbo-Kobayashi-Maskawa mixing 
of the quarks.
 So, Eq. (37) exhibits a boundary condition on the 
the renormalization group evolution for the effective theory
below $M_{\rm GUT}$, which we have assumed to 
be the MSSM. 
As we have demonstrated in a number of 
publications \cite{kmz,mondragon2,kmtz,kmtz2},
especially in \cite{kmoz2}, there are various 
supersymmetric GUTs with GYU in the 
third generation that can
 predict the bottom and top
quark masses in accordance with the experimental data.
This means that the top-bottom hierarchy 
could be
explained in these models,
 in a similar way as 
the hierarchy of the gauge couplings of the SM
can be explained if one assumes  the existence of a unifying
gauge symmetry at $M_{\rm GUT}$.

It is clear that the GYU scenario  is the most predictive scheme as far as
the mass of the top quark is concerned.
It may be worth recalling the predictions for $m_t$
of ordinary GUTs, in particular of supersymmetric $SU(5)$ and
$SO(10)$.  The MSSM with $SU(5)$ Yukawa boundary unification allows
$m_t$ to be anywhere in the interval between 100-200 GeV \cite{barger1}
for varying $\tan \beta$, which is now a free parameter.  Similarly,
the MSSM with $SO(10)$ Yukawa 
boundary conditions, {\em i.e.} $t-b-\tau$ Yukawa Unification gives
$m_t$ in the interval 160-200 GeV 
\cite{barbero,hall1,wright1,polonsky2}.

Clearly, to exclude or verify different GYU models,
 the experimental as well as theoretical uncertainties
have to be further reduced.
One of the largest theoretical uncertainties 
 for FUT, as we have seen,  results
from the not-yet-calculated threshold effects 
of the superheavy particles.
Since the structure of  the superheavy 
particles in FUT is basically fixed,
 it will be possible to
bring these threshold effects under control,
which will  reduce the uncertainty of 
the $M_t$ prediction ($5$ GeV) to $\sim  2$ GeV.
 We have been regarding $\delta^{\rm MSSM} M_t$ 
as unknown because we have no
sufficient information on the superpartner spectra.
Recently, however, it has been found that the principle
of finiteness \cite{jj} and also of  reduction of 
couplings \cite{future} can be applied to 
dimensionfull parameters,
e.g., soft breaking parameters, too. As a result, 
it becomes possible to predict the superpartner spectra 
to a certain extent and then to calculate $\delta^{\rm MSSM} M_t$.

It will be very interesting to find out in the coming years, as the
experimental accuracy of $m_t$ increases, if nature is kind enough to
verify our conjectured Gauge-Yukawa Unification.

\noindent{\bf Acknowledgements}
\vspace {0.5 cm}

J.K.  would like to thank the organizers for their
kind hospitality.

\end{document}